\documentclass[twocolumn]{aastex701}
\usepackage{bm}
\usepackage{amsmath}

\begin{document}

\title{Simulating AGN feedback in galaxy clusters with pre-existing turbulence}

\author{Jia-Lun Li}
\affiliation{Institute of Astronomy, National Tsing Hua University, Hsinchu 30013, Taiwan}
\email[show]{jialun.li@gapp.nthu.edu.tw (Jia-Lun Li)} 

\author{H.-Y. Karen Yang} 
\affiliation{Institute of Astronomy, National Tsing Hua University, Hsinchu 30013, Taiwan}
\affiliation{Physics Division, National Center for Theoretical Sciences, Taipei 106017, Taiwan}
\email[show]{\\ \hspace*{0.5cm} hyang@phys.nthu.edu.tw (H.-Y. Karen Yang)}

\begin{abstract}

Feedback from active galactic nuclei (AGN) is believed to play a significant role in suppressing cooling flows in cool-core (CC) clusters. Turbulence in the intracluster medium (ICM), which may be induced by AGN activity or pre-existing motions, has been proposed as a potential heating mechanism based on analysis of Chandra X-ray surface brightness fluctuations. However, subsequent simulation results have found the subdominant role of turbulence in heating the ICM. To investigate this discrepancy, we perform three-dimensional hydrodynamic simulations of a Perseus-like cluster including both AGN feedback and pre-existing turbulence, which is stirred to the observationally constrained level in the Perseus cluster. Our results indicate that, although the velocity field is dominated by the pre-existing turbulence, AGN heating through bubbles and shocks remains significant. More importantly, analysis of the velocity structure function and the energy power spectrum shows that the turbulent heating rate is smaller than the radiative cooling rate, especially in the cluster core. Our results offer insights relevant for recent XRISM observations and indicate that turbulent heating alone cannot offset radiative cooling in CC clusters.

\end{abstract}



\section{Introduction}

Cool-core (CC) clusters are galaxy clusters whose central gas has a radiative cooling time much shorter than the Hubble time. These clusters are expected to host massive gas inflows and exhibit high star formation rates (SFRs) \citep{Fabian94}. However, observational studies reveal the absence of such inflows, and SFRs are reduced by roughly an order of magnitude compared to predictions \citep{McNamara89, ODea08, McDonald11b, Hoffer12}. This is referred to as the ``cooling-flow problem.'' Due to the difference between the expectations and observations, it is suspected that there should be some heating mechanisms to balance radiative losses in cluster cores. Therefore, many heating mechanisms have been proposed to prevent catastrophic cooling. Notably, there is an observed correlation between the power of AGN jet-inflated cavities and the X-ray luminosity within the cores of CC clusters \citep{Best07, Mittal09, Birzan12}, supporting the belief that AGN feedback is an important mechanism for suppressing cooling flows in CC clusters \citep[see a recent review by][]{Bourne23}.

A crucial question regarding AGN feedback is how AGN jets heat the intracluster medium (ICM). One potential mechanism is turbulent dissipation. In this process, turbulence arises as a result of hydrodynamic instabilities as AGN jet-inflated bubbles are disrupted. This turbulence cascades to smaller scales, where its energy eventually dissipates as heat, contributing to the heating of the ICM and helping to balance cooling within the clusters. However, turbulence can arise not only from AGN activity but also from other sources, such as cluster mergers \citep{Ricker2001, ZuHone2010}, galaxy motions \citep{Ruszkowski11a}, and plasma instabilities \citep{McCourt11, Yang16a, Berlok2021}, all of which can generate turbulence within the ICM.

According to observations, turbulent heating has been shown to play a significant role in balancing radiative cooling. \cite{Zhuravleva14} derived the velocity power spectra of the hot ICM in the Perseus and Virgo clusters by making the assumption of a direct conversion between density fluctuations observed in the residual X-ray brightness map and the resulting turbulence-induced velocity fluctuations. By using the velocity power spectra they obtained, \cite{Zhuravleva14} calculated the turbulent dissipation rate, and demonstrated its capability to offset the radiative cooling rate. Their results suggest that turbulent heating alone is sufficient to balance radiative cooling in CC clusters. Similar results have been found in other CC clusters as well \citep{Zhuravleva2018}.

However, many simulation studies raise doubts regarding turbulent heating as the primary heating mechanism in galaxy clusters. \citet{Reynolds15} found that AGN is an inefficient driver of turbulence. By simulating self-regulated AGN feedback, \cite{Yang16b} and \cite{Li17} demonstrated that turbulent heating alone is insufficient to balance the radiative cooling. Additionally, \cite{Mohapatra_2019} found that, because of the constraint imposed by the presence of condensing multiphase gas from the ICM, which is consistent with observation data, the turbulent velocities required to balance cooling exceed the value of 164 $\pm$ 10 km s$^{-1}$ as measured in Perseus by the \textit{Hitomi} satellite \citep{Hitomi16}. Consequently, these simulation studies have challenged the hypothesis that turbulent heating is the dominant heating source in CC clusters. 

Moreover, the direct conversion between density fluctuations and velocity fluctuations which was applied in \cite{Zhuravleva14} is also being questioned. According to \cite{Mohapatra_2020}, strong stratification with a specific turbulent velocity can lead to larger density fluctuations, potentially resulting in an overestimation of the turbulent dissipation rate in the strongly stratified ICM. Furthermore, \cite{Wang_23} pointed out that strong stratification can suppress the turbulent cascade by channeling kinetic energy into buoyancy potential energy, thereby limiting the heating efficiency. In addition to turbulence, other processes such as contact discontinuities, shocks and AGN jet-inflated bubbles can also contribute to density fluctuations. Beyond this, \cite{Zhang22} showed that line-of-sight (LOS) velocities of H$_\alpha$ filaments in the model with rigid bubbles can also reproduce the observed X-ray-weighted gas velocity dispersion \citep{Hitomi16} without spatially homogeneous small-scale turbulence. 

In this work, we use three-dimensional (3D) hydrodynamic simulations incorporating both AGN feedback  and pre-existing turbulence to investigate the role of turbulent heating within a Perseus-like cluster. Our primary aim is to assess the potential contribution of turbulence to balancing radiative cooling, addressing discrepancies between previous simulation results and interpretations of observational data. To achieve this, our analysis first examines how the pre-existing turbulence affects AGN bubble evolution and its subsequent heating of the ICM. Building on this, we then compute the turbulent heating rate using two independent methods, analyzing the velocity structure function (VSF) and the energy power spectrum (\(E(k)\)), and compare it with the radiative cooling rate. We further compare the simulated LOS velocity dispersion with the data in light of the recent XRISM observations of the Perseus cluster \citep{XRISM25perseus}.

The paper is organized as follows. In Section~\ref{sec:2}, we describe the governing hydrodynamic equations, the simulation setup, and the Perseus-like cluster initial conditions. We also detail the modeling of pre-existing turbulence and AGN jet injection, as well as the two methods used to estimate the turbulent heating rate. In Section~\ref{sec:3}, we present the overall evolution of the cluster gas across different simulation runs (Section~\ref{sec:3.1}), compare the resulting velocity fields and velocity dispersions (Section~\ref{sec:3.2}), examine the entropy evolution and pressure perturbations (Section~\ref{sec:3.3}), analyze the impact of the jet on the energy spectrum (Section~\ref{sec:3.4}), and compare the turbulent heating and radiative cooling rates (Section~\ref{sec:3.5}).
In Section~\ref{sec:4}, we discuss the properties of the simulated turbulence, compare our estimated heating rates with previous simulations and observational inferences, explore the implications for XRISM observations, and outline the limitations of our simulation setup. Finally, Section~\ref{sec:5} summarizes our main conclusions.

\section{Methods}\label{sec:2}

We perform 3D hydrodynamic simulations of a Perseus-like galaxy cluster using the adaptive mesh refinement (AMR) code FLASH \citep{Flash, Dubey08}. The standard hydrodynamic equations, including source terms for AGN jet feedback and turbulence driving, are solved:
\begin{equation}
\frac{\partial \rho}{\partial t} + \nabla \cdot (\rho \mathbf{v}) = \dot{\rho}_{\rm jet},
\end{equation}
\begin{equation}
\frac{\partial (\rho \mathbf{v})}{\partial t} + \nabla \cdot (\rho \mathbf{v}\mathbf{v}) + \nabla P = \rho \mathbf{g}+ \mathbf{f}_{\rm jet} + \mathbf{f}_{\rm turb},
\end{equation}
\begin{equation}
\frac{\partial e}{\partial t} + \nabla \cdot [(e + P)\mathbf{v}] = \rho \mathbf{v} \cdot \mathbf{g}+\dot{e}_{\rm jet} + \dot{e}_{\rm turb},
\end{equation}
where \(\rho\) is the gas density, \(\mathbf{v}\) is the velocity, and \(\mathbf{g}\) is the gravitational field. The terms $\dot{\rho}_{\rm jet}$, $\mathbf{f}_{\rm jet}$, and $\dot{e}_{\rm jet}$ represent the mass, momentum, and energy injection from the AGN jets, respectively. The terms $\mathbf{f}_{\rm turb}$ and $\dot{e}_{\rm turb}$ represent the external forcing and corresponding energy injection associated with the isotropic stirring used to maintain pre-existing turbulence.
The total energy is given by \(e = 0.5\rho v^2 + e_i\), and the thermal pressure is \(P = (\gamma - 1)e_i\). 
An ideal gas equation of state with \(\gamma = 5/3\) is adopted. To isolate the effects of turbulent heating, radiative cooling is excluded from all simulations. Magnetic fields, thermal conduction, and physical viscosity are also omitted to avoid additional sources of complexity and to better characterize the fundamental hydrodynamic response of the ICM.  Viscosity in our simulations arises only through numerical dissipation. These idealized simulations therefore serve as a baseline for comparison with more realistic models incorporating cooling, magnetohydrodynamics (MHD), and anisotropic conduction in the future.

The simulation domain spans \(500\,\mathrm{kpc}\) on a side. The base grid (\texttt{lrefine\_min} = 6) has a spatial resolution of \(\sim 1.95\,\mathrm{kpc}\), while adaptive refinement up to level 7 yields a finest resolution of \(\sim 0.98\,\mathrm{kpc}\).  Refinement is applied within the central \(100\,\mathrm{kpc}\) region, ensuring higher spatial resolution in the cluster core. We have verified that the results are numerically converged at this resolution. A convergence test comparing a simulation using a uniform grid at level 6 against our AMR setup (refinement from level 6 to 7 in the central region) showed minimal differences in the key quantities of interest. Diode boundary conditions are applied on all sides of the domain. These are similar to outflow boundary conditions, but additionally prevent any inflow of gas into the simulation volume.\par

\subsection{Cluster Initial Conditions}
We adopt the same idealized CC cluster setup as in \citet{Yang16a}, with a temperature profile based on an analytical fit motivated by the observed X-ray surface brightness of the Perseus cluster:
\begin{equation}
T = T_0 \: \frac{1 + (r/r_0)^3}{2.3 + (r/r_0)^3} \left[ 1 + (r/r_1)^2 \right]^{-0.32},
\end{equation}
where \( T_0 = 7\,\mathrm{keV} \), \( r_0 = 71\,\mathrm{kpc} \), and \( r_1 = 380\,\mathrm{kpc} \). The gas density profile is obtained by assuming hydrostatic equilibrium in a static Navarro–Frenk–White (NFW; \citealt{Navarro96}) gravitational potential:
\begin{equation}
\Phi(r) = -\frac{GM}{r}  \frac{\ln(1 + r/r_s)}{\ln(1 + c) - c/(1 + c)},
\end{equation}
where \( M \) is the cluster virial mass, \( r_{\rm s} \equiv r_{\rm vir}/c \) is the scale radius, \( r_{\rm vir} \) is the virial radius, and \( c \) is the concentration parameter. In all simulations, these parameters are set to
\( M = 8.5 \times 10^{14} \, M_{\odot} \), \( r_{\rm vir} = 2.440\,\mathrm{Mpc} \), and \( c = 6.81 \).

\subsection{Turbulence Driving}\label{sec:driving}
To model turbulence in the ICM, we utilize the Stir module in the FLASH code, which employs a spectral forcing scheme in Fourier space that generates statistically stationary velocity fields \citep{refId0,ESWARAN1988257}. This scheme is based on an Ornstein–Uhlenbeck (OU) random process, which generates a forcing acceleration field that evolves smoothly over time with a finite autocorrelation time. The acceleration field has zero mean and a constant root-mean-square amplitude, characteristics that allow us to model the turbulent driving with realistic temporal correlations. The acceleration for each Fourier mode is then updated according to:
\begin{equation}
{a}_{n+1} = f\, {a}_n + \sigma_a \sqrt{1 - f^2}\, {G}_n,
\end{equation}
where \({a}_n\) is the acceleration at the current timestep, \(f = \exp(-\Delta t/\tau_d)\) is the exponential damping factor defined by the correlation time \(\tau_d\) and timestep \(\Delta t\), \(\sigma_a\) is the standard deviation of the forcing amplitude related to the energy injection rate \(\epsilon\) through \(\sigma_a^2 = \epsilon / \tau_d\), and \({G}_n\) is the Gaussian random variable. The factor \(\sqrt{1 - f^2}\) ensures that the stochastic contribution is properly scaled so the total variance of the acceleration remains constant. To preferentially excite vortical motions and mimic a Kolmogorov-like turbulent cascade, we restrict the forcing to solenoidal modes only.

To ensure that our simulations are consistent with observational constraints of the Perseus cluster from \citet{Hitomi16}, which report a characteristic LOS velocity dispersion \(\sigma_{\rm LOS} \sim 164~\mathrm{km\,s^{-1}}\), we adjust the relevant simulation parameters by comparing the domain-averaged simulated \(\sigma_{\rm LOS}\) to this value (see Table~\ref{table:1}). The turbulent driving is applied over a wavenumber range \(k_{\rm min} \leq k \leq k_{\rm max}\), corresponding to physical scales from \(L_{\rm max} = 2\pi/k_{\rm min} \approx 250~\mathrm{kpc}\) down to \(L_{\rm min} = 2\pi/k_{\rm max} \approx 100~\mathrm{kpc}\). Our simulation setup with turbulent driving is similar to \citet{Ruszkowski10}. For reference, they listed the 3D velocity dispersion, which is larger than the LOS value by a factor of \(\sqrt{3}\) for isotropic turbulence.

\setlength{\tabcolsep}{4pt}
\begin{deluxetable}{ccccc}[t!]
\tabletypesize{\footnotesize}
\tablecaption{Parameters used for turbulence driving in the simulations.\label{table:1}}
\tablewidth{0pt}
\tablehead{
    \colhead{\(k_{\rm min}\)} & 
    \colhead{\(k_{\rm max}\)} & 
    \colhead{\(\tau_{\rm d}\)} & 
    \colhead{\(\epsilon\)} & 
    \colhead{\(\sigma_{\rm LOS}\)} \\
    \colhead{(cm\(^{-1}\))} & 
    \colhead{(cm\(^{-1}\))} & 
    \colhead{(s)} & 
    \colhead{(cm\(^2\) s\(^{-3}\))} & 
    \colhead{(km s\(^{-1}\))}
}
\startdata
$8.14 \times 10^{-24}$ & $2.04 \times 10^{-23}$ & $8.5 \times 10^{15}$ & $4 \times 10^{-5}$ & $\sim 185$ \\
\enddata
\end{deluxetable}

\begin{figure}
    \centering
    \includegraphics[width=\columnwidth]{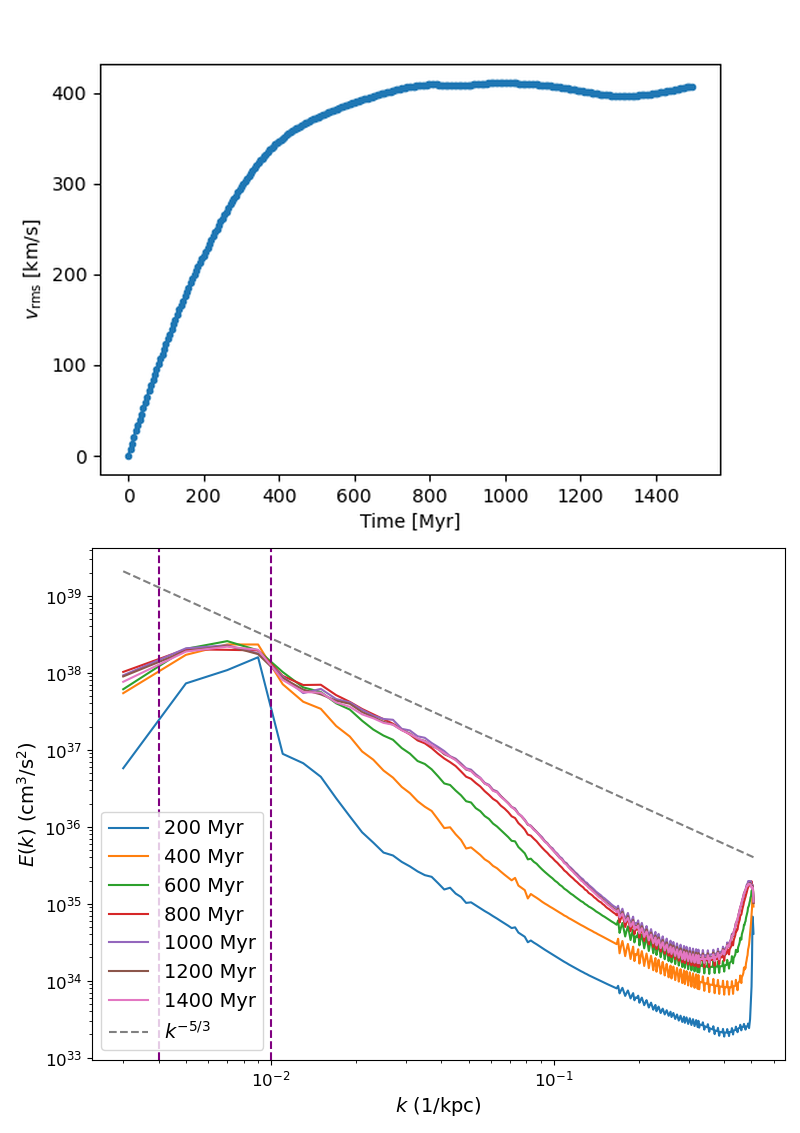}
    \caption{Development of the pre-existing turbulence. Upper panel: Time evolution of the 3D root-mean-square (RMS) turbulent velocity, showing that the system with pre-existing turbulence reaches a steady state after $\sim 800~\mathrm{Myr}$. Lower panel: Time evolution of the kinetic energy spectrum, $E(k)$, shown from $t=200$ to $1400~\mathrm{Myr}$ with an interval of $200~\mathrm{Myr}$. The convergence of the spectra to a stable profile confirms that the turbulent cascade is fully established prior to the jet injection. The slight uptick in power at $k \gtrsim 0.4\,{\rm kpc}^{-1}$ is a numerical artifact introduced by the resampling (resolution-conversion) procedure required to perform the FFT on a uniform grid. This feature arises near the effective resolution limit where the spectrum transitions into the numerical dissipation regime, and has no impact on the inertial-range scaling or the physical analysis presented in this work.}
    \label{fig1}
\end{figure}

\subsection{Simulation Models and AGN Feedback}
To investigate the individual and combined roles of turbulence and AGN feedback in ICM heating, we perform three simulations with distinct setups. The \textbf{Turb+Jet} run includes both pre-existing turbulence and a single episode of bipolar AGN jet activity. The \textbf{TurbOnly} run includes only pre-existing turbulence, while the \textbf{JetOnly} run includes only the jet. These controlled setups allow us to isolate and quantify the separate and joint contributions of turbulence and AGN jets to the heating of the ICM.

The AGN jet in the \textbf{Turb+Jet} and \textbf{JetOnly} runs is purely kinetic, with a power of \(5 \times 10^{45}\,\mathrm{erg\,s^{-1}}\) and a duration of \(10\,\mathrm{Myr}\), consistent with the average jet power from previous self-regulated feedback simulations of a Perseus-like cluster \citep{Yang16b, Dunn04}. The bipolar jets are launched after the turbulence reaches a quasi-steady state and are injected along the \(\pm z\)-axis within a cylindrical region of radius \(2\,\mathrm{kpc}\) and height \(4\,\mathrm{kpc}\). No precession is included in this setup. This configuration provides a clean baseline to examine how the jets interact with the ICM in the presence or absence of pre-existing turbulence.

\subsection{Estimating Turbulent Heating Rates}\label{sec:heating}
We estimate the turbulent heating rate in our simulations using two complementary methods: VSF and \(E(k)\) \citep{Pope2000}.

\subsubsection{Velocity Structure Functions}
The VSF quantifies how velocity differences between two points in a turbulent flow depend on their spatial separation. It characterizes the distribution of energy across different scales and is especially useful in identifying scaling behaviors within the inertial range.

Specifically, the total second-order velocity structure function, \(VSF_2\), calculates the average squared magnitude of the velocity difference between two points separated by a distance \(\ell\), providing a statistical measure of the total intensity of turbulent fluctuations at that scale:
\begin{equation}
VSF_2(\ell) = \left\langle \left| \mathbf{v}(\mathbf{x} + \boldsymbol{\ell}) - \mathbf{v}(\mathbf{x}) \right|^2 \right\rangle,
\label{eq:total_vsf2}
\end{equation}
where \(\boldsymbol{\ell}\) is the separation vector, \(\mathbf{v} = (v_x, v_y, v_z)\) is the 3D velocity field, and the angle brackets denote spatial averaging. This expression captures the total velocity difference across all directions for a given scale \(\ell = |\boldsymbol{\ell}|\).

In turbulence theory, the velocity difference is often decomposed into components parallel and perpendicular to the separation vector. The longitudinal second-order velocity structure function, \(D_{\rm LL}(\ell)\), measures the variance of the velocity difference component parallel to \(\boldsymbol{\ell}\):
\begin{equation}
D_{\rm LL}(\ell) = \left\langle \left( \left[ \mathbf{v}(\mathbf{x} + \boldsymbol{\ell}) - \mathbf{v}(\mathbf{x}) \right] \cdot \hat{\boldsymbol{\ell}} \right)^2 \right\rangle,
\label{eq:longitudinal_vsf2}
\end{equation}
where \(\hat{\boldsymbol{\ell}} = \boldsymbol{\ell}/\ell\). Similarly, the transverse structure function, \(D_{\rm NN}(\ell)\), quantifies the variance of the velocity difference components perpendicular to \(\boldsymbol{\ell}\). Assuming statistical isotropy, the total \(VSF_2\) is related to the longitudinal and transverse components by \(VSF_2(\ell) = D_{\rm LL}(\ell) + 2D_{\rm NN}(\ell)\). To ensure consistency with the standard Kolmogorov formulation and to derive the dissipation rate using established constants, our analysis focuses on \(D_{\rm LL}(\ell)\) as defined in Eq.~\eqref{eq:longitudinal_vsf2}.

According to Kolmogorov's theory for incompressible turbulence, energy cascades from large to small scales, leading to a characteristic scaling law within the inertial range \citep{Kolmogorov_1941}. In this framework, \(D_{\rm LL}(\ell)\) follows the scaling relation \(D_{\rm LL}(\ell) \propto \ell^{2/3}\) \citep{Pope2000}. The turbulent dissipation rate per unit mass, \(\varepsilon\), can be estimated by fitting Eq.~\eqref{eq:dll} to the measured \(D_{\rm LL}(\ell)\) within the inertial range:
\begin{equation}
D_{\rm LL}(\ell) = C_2 \, \varepsilon^{2/3} \ell^{2/3},
\label{eq:dll}
\end{equation}
where \(C_2 \approx 2.0\) is the dimensionless Kolmogorov constant \citep{Saddoughi1994}. We note that the dissipation rate \(\varepsilon\) represents the actual rate of turbulent energy dissipation, which is distinct from the injection rate \(\epsilon\) that denotes the imposed energy input in the simulation.

\begin{figure*}
    \centering
    \includegraphics[width=1\textwidth]{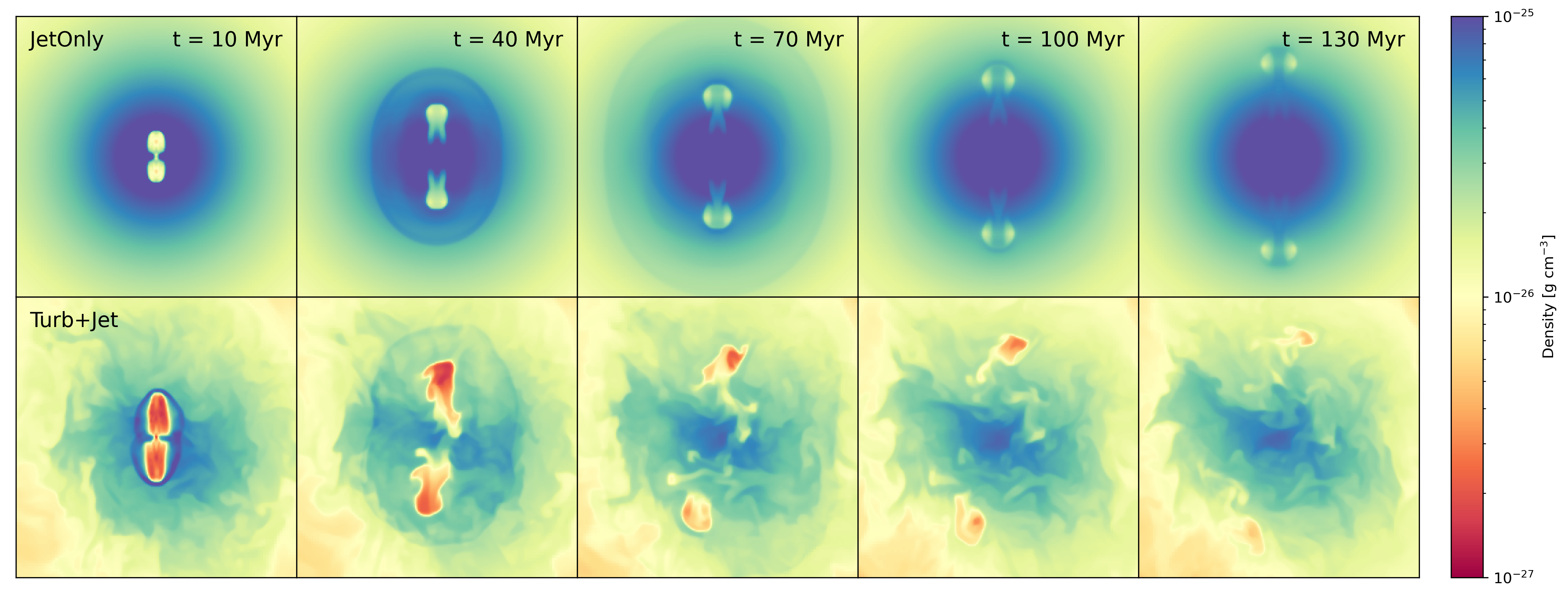}
    \caption{Time evolution of the gas density slices at the x = 0 plane for \textbf{JetOnly} (top row) and \textbf{Turb+Jet} (bottom row). Columns correspond to snapshots taken at \(t = 10\), 40, 70, 100, and 130~Myr after jet injection. Each slice is 200~kpc on a side. In \textbf{Turb+Jet}, the jets propagate faster and the cluster core is less centrally concentrated due to pre-existing turbulence, which also disrupts the bubbles more quickly compared to \textbf{JetOnly}.
    }
    \label{fig2}
\end{figure*}

\begin{figure*}
    \centering
    \includegraphics[width=0.75\textwidth]{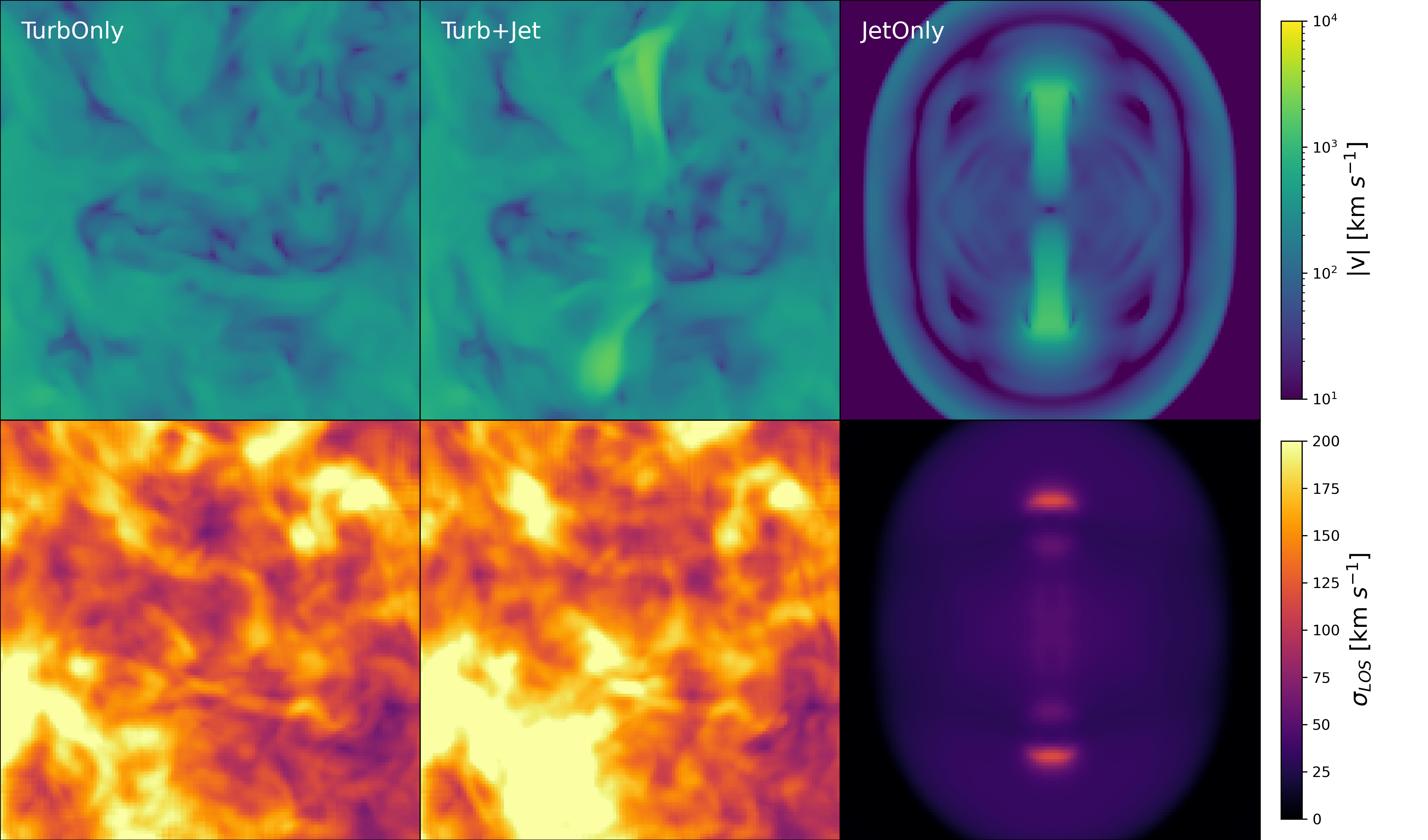}
    \caption{Columns from left to right correspond to \textbf{TurbOnly}, \textbf{Turb+Jet}, and \textbf{JetOnly}, shown at the same timestep, 50~Myr after the jet injection. The first row shows thin projections (4~kpc) of the velocity magnitude ($|{\boldsymbol v}|$) weighted by gas density. The second row shows the line-of-sight velocity dispersion ($\sigma_{\rm LOS}$) weighted by X-ray emissivity.Each panel spans 132~kpc on a side. Both $\lvert v \rvert$ and $\sigma_{\rm LOS}$ are dominated by pre-existing turbulence, indicating that the jet has a minor influence on the overall velocity field.
    }
    \label{fig3}
\end{figure*}

\subsubsection{Energy Power Spectrum}

Alternatively, we compute the 3D \(E(k)\) from the velocity field using a fast Fourier transform (FFT). In the inertial range, Kolmogorov theory predicts that \(E(k)\) scales as:
\begin{equation}
E(k) = C_K\, \varepsilon^{2/3} k^{-5/3},
\label{eq:ek}
\end{equation}
where \(\varepsilon\) is the turbulent dissipation rate per unit mass, and \(C_K \approx 1.65\) is the Kolmogorov constant \citep{10.1063/1.868656, 10.1063/1.1539855}. We estimate \(\varepsilon\) by fitting the inertial-range portion of the spectrum to the expected scaling relation.

\section{Results}\label{sec:3}

\subsection{Overall Evolution}\label{sec:3.1}
In the cases with pre-existing turbulence, homogeneous turbulent fields develop during the initial growth phase, leading to a gradual increase of the root-mean-square (RMS) turbulent velocity \(v_{\rm rms}\). This build-up phase lasts for roughly 800 Myr, which is comparable to the eddy turnover time \(t_{\rm eddy}\) at the driving scale and approximately equal to the autocorrelation time \(\tau_{\rm d}\) of the OU forcing. After this period, \(v_{\rm rms}\) saturates at a nearly constant value, indicating that a statistical steady state has been established where energy injection and dissipation are balanced (Fig.~\ref{fig1}). For the subsequent single-jet injection, we select a snapshot at approximately 1255 Myr as the injection time, well after the turbulence has reached this quasi-steady regime to ensure that the system is fully equilibrated.

Fig.~\ref{fig2} shows the density evolution following the jet injection for both \textbf{JetOnly} and \textbf{Turb+Jet}. In \textbf{Turb+Jet}, the jets propagate faster than in \textbf{JetOnly}. This is because, prior to the jet injection, the pre-existing turbulence lowers the central density, preventing a strong density peak from forming and leaving the core region more diffuse. The turbulent motions also disrupt the jet-inflated bubbles more rapidly, and by roughly 130~Myr after the injection, the bubbles are no longer visible in the density slices.

\subsection{Velocity Field}\label{sec:3.2}
\begin{figure}
    \centering
    \includegraphics[width=\columnwidth]{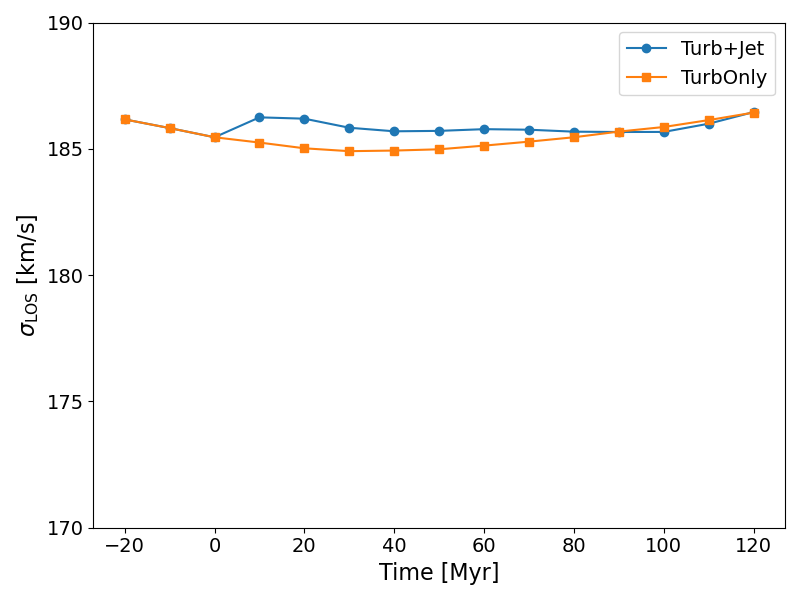}
    \caption{Time evolution of $\sigma_{\rm LOS}$ averaged over the entire simulation domain for \textbf{Turb+Jet} and \textbf{TurbOnly} as a function of the time since jet injection. The overall $\sigma_{\rm LOS}$ is similar in both cases, indicating that the jet has a minor impact and that the velocity field is dominated by pre-existing turbulence.
    }
    \label{fig4}
\end{figure}

In Fig.~\ref{fig3}, we present the velocity magnitude (\(|v|\)) and \(\sigma_{\rm LOS}\) 50 Myr after the jet injection for the three cases. The dispersion is defined as
\[
\sigma_{\rm LOS} = \sqrt{\langle v_x^2 \rangle - \langle v_x \rangle^2}\,,
\]
where \(v_x\) is the velocity along the LOS, and the brackets represent emissivity-weighted averages. A higher dispersion indicates stronger turbulence or random motions. For \textbf{TurbOnly}, where no jet is injected, we show the same time step for comparison. In this case, \(\sigma_{\rm LOS}\) is $\sim 185~{\rm km\,s^{-1}}$, consistent with the expected turbulence level described in Section~\ref{sec:driving}. In \textbf{JetOnly}, the velocity field is primarily shaped by the jet, producing a more ordered bipolar structure along the jet axis. In \textbf{Turb+Jet}, this coherent bipolar structure is still evident in \(|v|\), superimposed on the strong turbulent background. However, the \(\sigma_{\rm LOS}\) map, along with the overall velocity statistics, appears dominated by the pre-existing turbulence, closely resembling \textbf{TurbOnly}. Furthermore, Fig.~\ref{fig4} illustrates the time evolution of \(\sigma_{\rm LOS}\) after jet injection comparing \textbf{TurbOnly} and \textbf{Turb+Jet}. The overall velocity dispersion is similar in both cases, confirming that the jet has a minor influence on the global \(\sigma_{\rm LOS}\) value, despite its visible impact on \(|v|\) seen in Fig.~\ref{fig3}.

%In \textbf{Turb+Jet} and \textbf{TurbOnly}, both \(\lvert v \rvert\) and \(\sigma_{\rm LOS}\) are dominated by the pre-existing turbulence, which overwhelms any coherent large-scale flows driven by the jet. By contrast, in \textbf{JetOnly} the velocity field is primarily shaped by the jet, producing a more ordered bipolar structure along the jet axis. As a result, compared to the pre-existing turbulence, the jet appears to have little impact on the velocity field. Furthermore, Fig.~\ref{fig4} illustrates the time evolution of \(\sigma_{\rm LOS}\) after jet injection comparing \textbf{TurbOnly} and \textbf{Turb+Jet}.
%The overall velocity dispersion is similar in both cases, indicating that the jet has a minor influence on \(\sigma_{\rm LOS}\). This is consistent with the dominance of pre-existing turbulence seen in Fig.~\ref{fig3}.

\begin{figure*}
    \centering
    \includegraphics[width=1\textwidth]{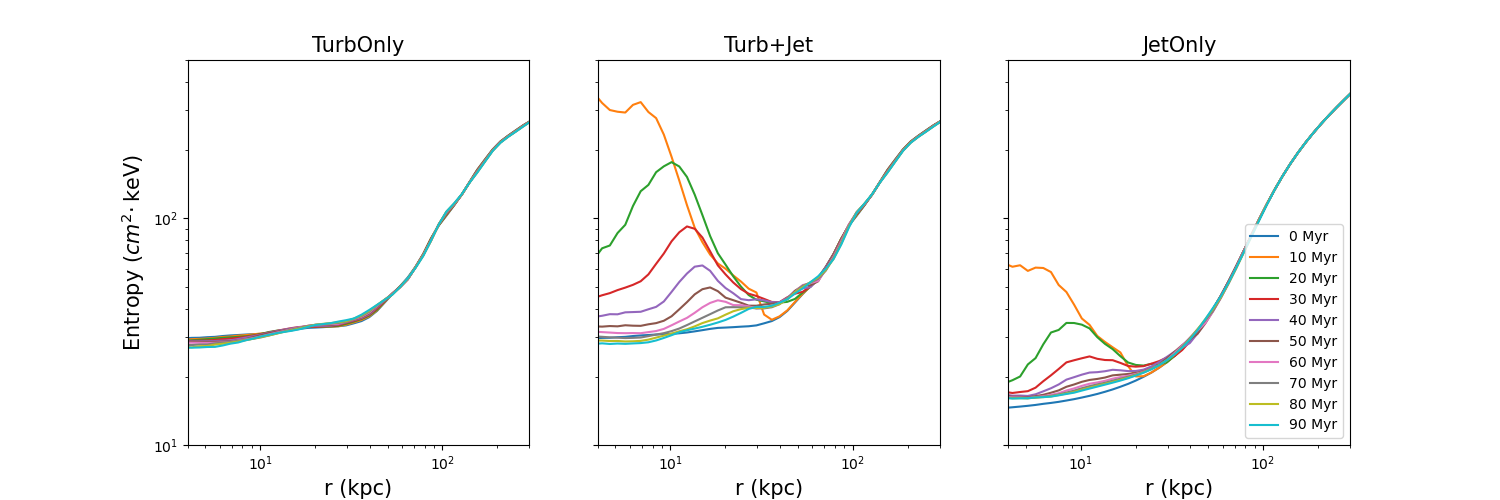}
    \caption{Entropy profiles as a function of radius for \textbf{TurbOnly}, \textbf{Turb+Jet}, and \textbf{JetOnly}, shown at the same timesteps after jet injection. Different lines correspond to different timesteps. The entropy increase is mainly due to the jet, with turbulence having little effect, as shown by the similar profiles of \textbf{Turb+Jet} and \textbf{JetOnly}, while \textbf{TurbOnly} remains nearly constant.
    }
    \label{fig5}
\end{figure*}

\begin{figure*}
    \centering
    \includegraphics[width=0.8\textwidth]{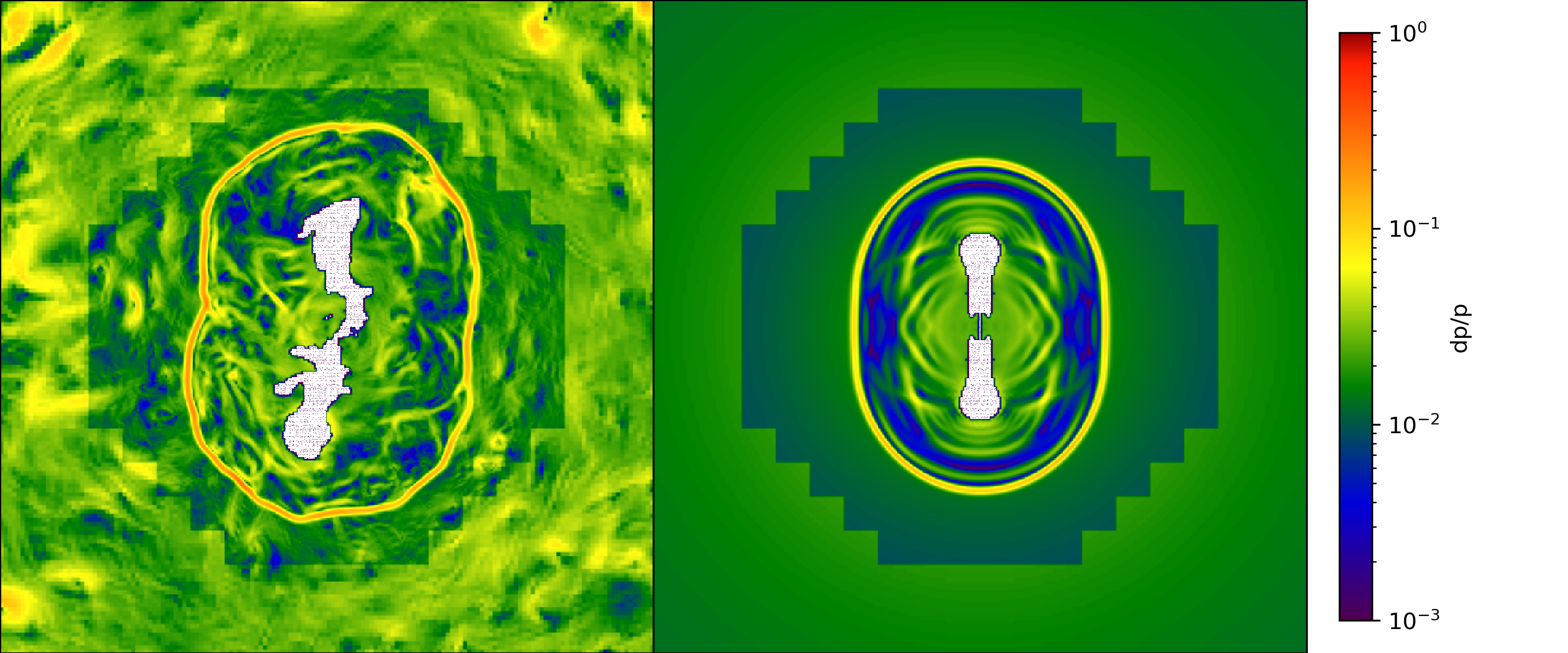}
    \caption{Slices of $\Delta P / P$ for \textbf{Turb+Jet} (left) and \textbf{JetOnly} (right), shown at 50 Myr after the jet injection. Each image spans 300 kpc on a side. Both sound waves and weak shocks are present, and the similarity between the two simulations indicates that jet-driven heating remains effective despite the dominance of pre-existing turbulence in the velocity field.
    }
    \label{fig6}
\end{figure*}

\subsection{Entropy Evolution and Pressure Perturbations}\label{sec:3.3}
In Fig.~\ref{fig5}, we present the entropy profiles for the three cases. Each colored curve represents a different time step after the jet injection. Although \textbf{TurbOnly} does not include any jet injection, its profiles are shown at the same time steps for comparison. The entropy is defined as
\[
K = \frac{k_{\mathrm{B}} T}{n_{\mathrm{e}}^{2/3}},
\]
where \( k_{\mathrm{B}} \) is the Boltzmann constant, \( T \) is the temperature, and \( n_{\mathrm{e}} = \rho/(\mu_{\rm e} m_{\rm p}) \) is the electron number density with \( \mu_{\rm e} = 1.17 \), typical for the ICM. An increase in entropy reflects energy input into the gas and corresponds to heating. As seen in the figure, \textbf{Turb+Jet} and \textbf{JetOnly} show nearly identical entropy evolution. The only difference between these two cases is the presence of turbulence, suggesting that turbulence has little effect on the heating, and that most of the entropy increase comes from the jet injection. This finding is further supported by \textbf{TurbOnly}, which shows almost no change in entropy over time, indicating that turbulence alone does not significantly contribute to heating. Note that \textbf{JetOnly} has a lower initial entropy than the turbulent runs, because the absence of turbulence keeps its central gas density high, while turbulence in the other runs reduces the density near the center, increasing the entropy.

Fig.~\ref{fig6} shows the relative pressure variations, \(\Delta P / P\), which reflect both small-amplitude sound waves and weak shocks. This allows us to identify these features and to estimate the Mach number of the corresponding gas motions. For small-amplitude sound waves (\(M_{\rm wave} \equiv \delta v / c_s \ll 1\)), the relative pressure perturbation is approximately linear with the wave Mach number:
\[
\frac{\delta P}{P} \sim \gamma M_{\rm wave},
\]
where \(\gamma = 5/3\) is the adiabatic index of the gas. For weak shocks (\(M_{\rm shock} \gtrsim 1\)), the relative pressure jump is given by
\[
\frac{\Delta P}{P} \equiv \frac{P_2 - P_1}{P_1} = \frac{2 \gamma}{\gamma + 1} (M_{\rm shock}^2 - 1),
\]
where \(P_1\) and \(P_2\) are the pre-shock and post-shock pressures, respectively.

We exclude cold gas (\(T \le 5 \times 10^{5}~\mathrm{K}\)) to focus on the hot gas component, which dominates the X-ray emission. The structures associated with jet injection, as seen in \textbf{JetOnly}, are still visible in \textbf{Turb+Jet}. In the hot ICM, the observed \(\Delta P / P\) spans both small and moderate values, with the smaller perturbations corresponding to subsonic motions and the larger perturbations, where \(\Delta P / P > 0.2\), indicating weak shocks \citep{Yang16b}. The similarity of \(\Delta P / P\) values associated with shocks and sound waves between the two simulations shows that, even though the overall velocity field is dominated by pre-existing turbulence, heating by the AGN jets remains effective. This again supports the discussion around Fig.~\ref{fig5} above.

\subsection{Assessing the Jet's Contribution to Turbulence}\label{sec:3.4}
Our previous analyses of the velocity field hint at the complex interplay between the jet and the pre-existing turbulence. The time evolution of $\sigma_{\rm LOS}$ when averaged over the entire simulation domain (Fig.~\ref{fig4}) shows a surprisingly minor difference between \textbf{Turb+Jet} and \textbf{TurbOnly}. Similarly, the velocity maps at $t=50 \text{ Myr}$ after jet injection (Fig.~\ref{fig3}) appear visually dominated by the strong background turbulence, masking the jet's coherent structure in the $\sigma_{\rm LOS}$ map. These domain-averaged and later-time views, however, may obscure the more dynamic interaction occurring within the cluster core at early times. It is therefore important to investigate the true nature of this early-time interaction to determine whether the jet genuinely feeds the stochastic turbulent cascade, or merely drives transient, coherent flows that disappear quickly.

\begin{figure*}
    \centering
    \includegraphics[width=\textwidth]{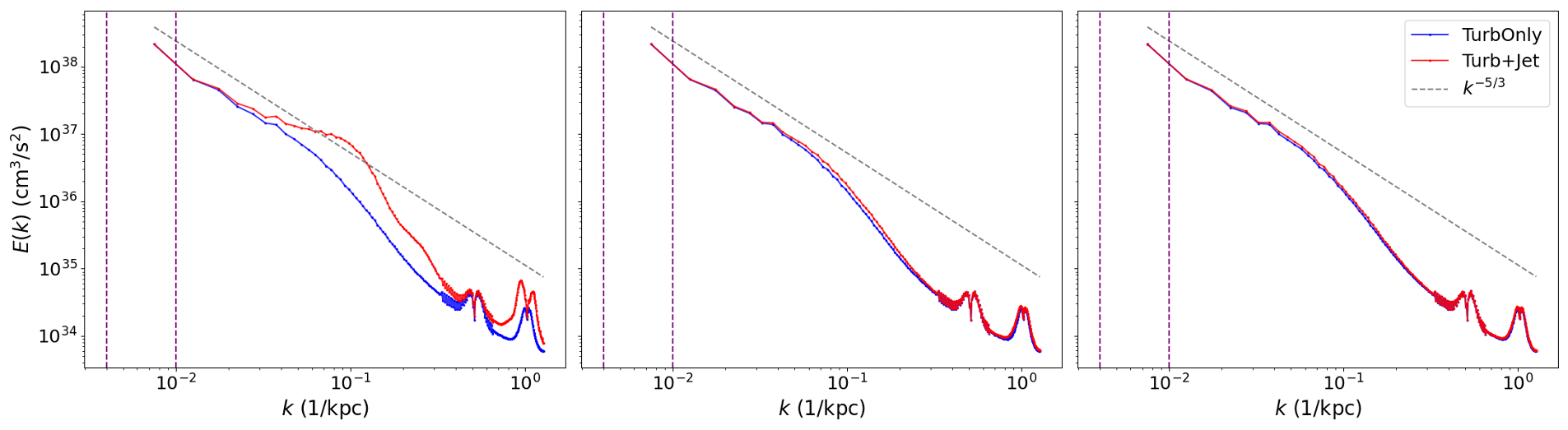}
    \caption{Energy spectra $E(k)$ as a function of wavenumber $k$ $(k=1/\ell)$ for \textbf{TurbOnly} and \textbf{Turb+Jet} at $t=10$, 20, and 30 Myr after jet injection, respectively from left to right. Each panel shows the spectra computed within a central region $200\,\mathrm{kpc}$ on a side.  The jet produces a noticeable but short-lived modification to the spectral shape. This effect fades quickly, after which the spectra in both runs appear similar, indicating that these transient, jet-induced motions do not effectively feed the turbulent cascade.
    }
    \label{fig7}
\end{figure*}

A single-value statistic like $\sigma_{\rm LOS}$ is insufficient to distinguish between these two scenarios. We therefore turn to $E(k)$, which can precisely map how the kinetic energy is distributed across different scales. In Fig.~\ref{fig7}, we present $E(k)$ computed specifically within a central region $200\,\mathrm{kpc}$ on a side at $t=10, 20$, and $30 \text{ Myr}$ after the jet injection. This figure shows that at $t=10 \text{ Myr}$, the \textbf{Turb+Jet} spectrum exhibits a bump compared to \textbf{TurbOnly}. However, this feature is transient. By $t=20 \text{ Myr}$, this excess power has substantially diminished, and by $t=30 \text{ Myr}$, the spectra of the two runs are nearly indistinguishable. 

This rapid disappearance indicates that the impact of the jet on the power spectrum is short-lived, and critically, that the energy injected by the jet at large scales does not cascade down to smaller scales, in contrast to the spectrum of driven turbulence shown in Fig.~\ref{fig1}. The evidence from $E(k)$ demonstrates that the gas motions driven by the jet are not turbulence, in the sense of a self-sustaining, stochastic cascade. Instead, they are best described as large-scale, coherent flows (often referred to as bulk motions) that are short-lived. These motions appear to dissipate, likely through mechanisms such as weak shocks or simple advection, without breaking down into smaller eddies and feeding the turbulent cascade. Therefore, within the context of our single-injection setup, the AGN jet appears to be a relatively inefficient driver of sustained turbulence, consistent with previous findings \citep{Reynolds15, Yang16b}. This implies that the turbulent kinetic energy in the cluster core is still dominated by the pre-existing background turbulence. In the following section, we will quantify the heating associated with this turbulence.

\subsection{Turbulent Dissipation Rate}\label{sec:3.5}
We estimate the turbulent heating rate using two independent approaches described in Section~\ref{sec:heating}, namely the VSF and \(E(k)\) methods. Both approaches are based on the Kolmogorov turbulence framework and provide estimates of \(\varepsilon\), characterizing turbulence in real and Fourier space, respectively.

\begin{figure}
    \centering
    \includegraphics[width=\columnwidth]{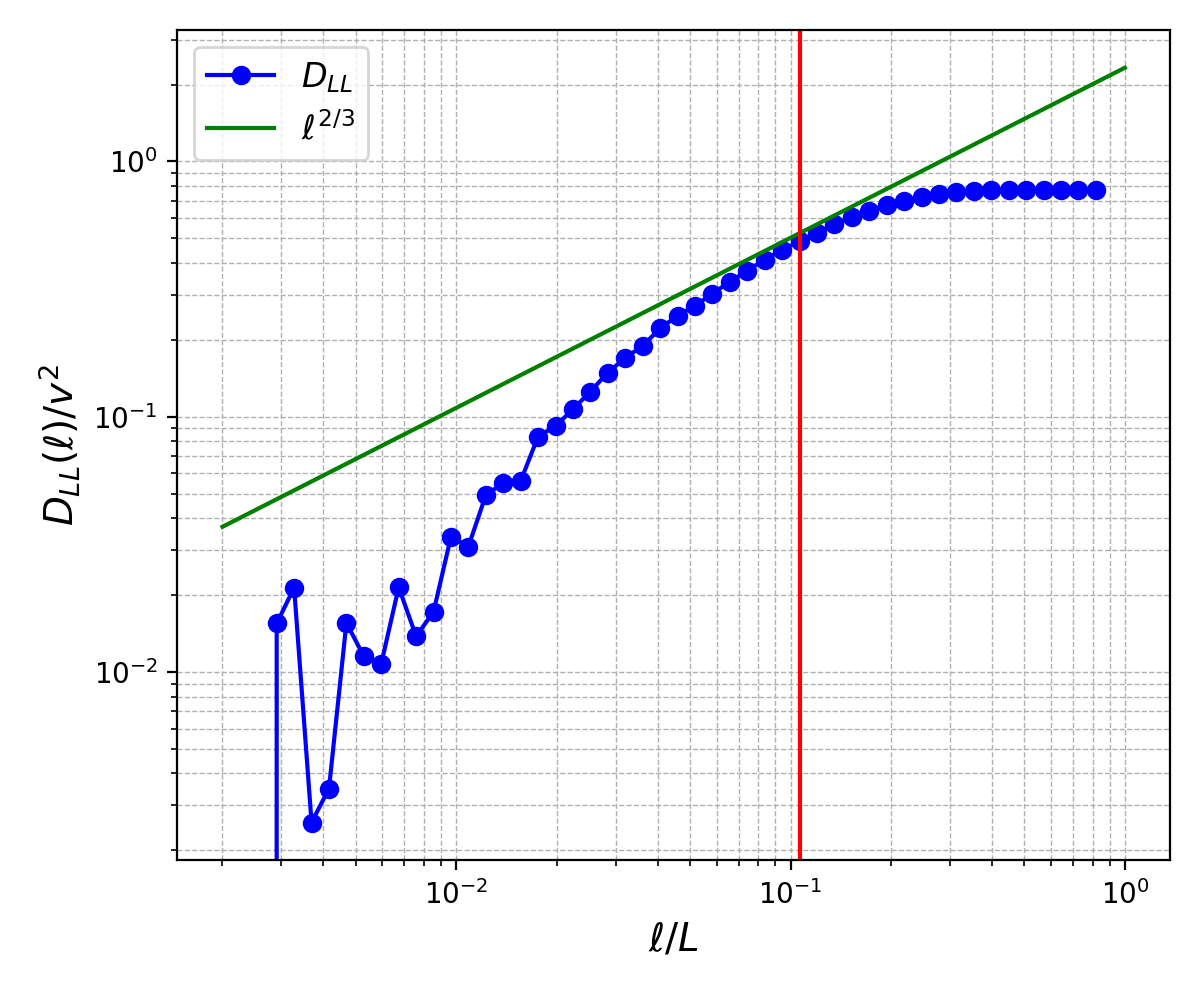}
    \caption{$D_{\rm LL}$ normalized by the volume-averaged squared velocity $\langle v^2 \rangle = 1.35\times10^{15}\ \mathrm{cm}^2\ \mathrm{s}^{-2}$ is shown as a function of the normalized separation $\ell/L$, where $L = 500~\mathrm{kpc}$ is the simulation box size. The green line indicates the expected Kolmogorov scaling, $D_{\rm LL} \propto \ell^{2/3}$. The vertical red line marks the characteristic scale $\ell_{\rm VSF} \approx 53.1~\mathrm{kpc}$ identified via the tangent point, which is used to estimate the turbulent dissipation rate $\varepsilon$.
    }
    \label{fig8}
\end{figure}

All results shown in this section are computed at $\sim1305$~Myr. The VSF and $E(k)$ calculations reported below are performed on the \textbf{TurbOnly} run, and we have verified that the \textbf{Turb+Jet} case at the same snapshot (corresponding to $\sim50$~Myr after its jet injection) exhibits very similar behaviour with no significant differences in either the VSF or $E(k)$ profiles due to the dominance of pre-existing turbulence in the stochastic components of the velocity field as described above. The resulting $D_{\rm LL}(\ell)$ is shown in Fig.~\ref{fig8}. At large scales, the VSF flattens because the velocity fluctuations at widely separated points become statistically uncorrelated. As the correlation approaches zero, $D_{\rm LL}(\ell)$ asymptotically approaches a constant saturation value. At small scales, the VSF steepens due to numerical dissipation \citep{Feder10,Simonte2022}, which smooths out small-scale motions. Only the intermediate range, approximately 20--100~kpc, follows the expected $\ell^{2/3}$ scaling predicted by Kolmogorov turbulence, corresponding to the inertial range where turbulent energy cascades efficiently from large to small scales.

\begin{figure}
    \centering
    \includegraphics[width=\columnwidth]{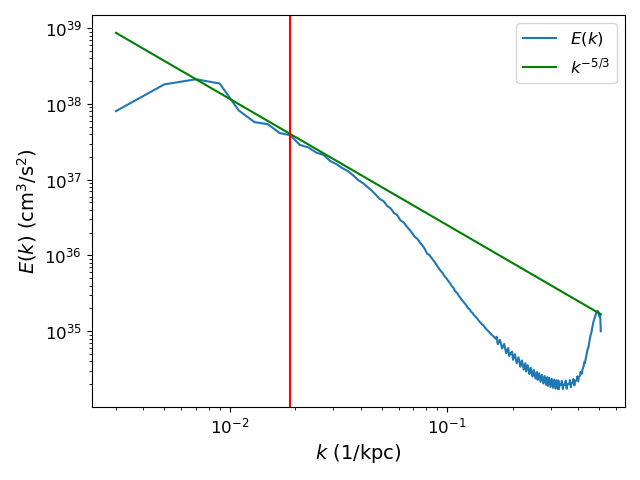}
    \caption{$E(k)$ as a function of wavenumber $k$ $(k=1/\ell)$. The green line indicates the expected Kolmogorov scaling, $E(k) \propto k^{-5/3}$, in the inertial range. The vertical red line marks the characteristic scale $\ell_{\rm Ek} \approx 52.6~\mathrm{kpc}$ identified via the tangent point, which is used to estimate the turbulent dissipation rate $\varepsilon$.
    }
    \label{fig9}
\end{figure}

The energy spectrum \(E(k)\), shown in Fig.~\ref{fig9}, describes how kinetic energy is distributed among different wavenumbers \(k\). The spectrum declines at both large and small scales. At low \(k\) (large scales, \(k \lesssim 0.004\)), approaching the fundamental mode of the simulation box and scales larger than the driving range, the spectrum falls off due to the lack of direct energy input combined with poor statistical sampling from the limited number of large-scale Fourier modes. At high \(k\) (small scales, \(k \gtrsim 0.05\)), the spectrum drops as the cascade approaches the dissipation range due to numerical dissipation. Between these two regimes lies the inertial range, which is characterized by the classical Kolmogorov scaling \(E(k) \propto k^{-5/3}\).

\begin{figure}
    \centering
    \includegraphics[width=\columnwidth]{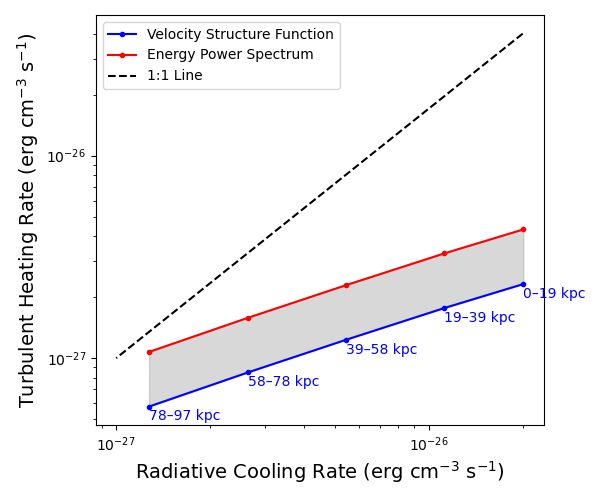}
    \caption{Turbulent heating rate $Q_{\rm turb}$ compared to the radiative cooling rate $Q_{\rm cool}$ in different radial shells. Solid lines show the heating rate estimates from the VSF (blue) and \(E(k)\) (red) methods at their characteristic scales, $\ell_{\rm VSF} \approx 53.1~\mathrm{kpc}$ and $\ell_{\rm Ek} \approx 52.6~\mathrm{kpc}$. Shaded bands indicate the uncertainty due to methodological differences. The black dashed line marks the $1{:}1$ relation, indicating where the turbulent heating rate equals the radiative cooling rate. Overall, the turbulent heating rate is smaller than the radiative cooling rate in the cluster center, while it becomes more comparable at larger radii.
    }
    \label{fig10}
\end{figure}

From the VSF and \(E(k)\) profiles shown in Figs.~\ref{fig8} and \ref{fig9}, we identify characteristic scales within the inertial range to estimate the turbulent dissipation rate $\varepsilon$. These scales are chosen by identifying points tangent to the theoretical Kolmogorov power laws ($D_{\rm LL} \propto \ell^{2/3}$ and $E(k) \propto k^{-5/3}$), corresponding to the maximum inferred energy transfer rate. As indicated in the figures, these points occur around $\ell_{\rm VSF} \approx 53.1$~kpc for the VSF and $\ell_{\rm Ek} \approx 52.6$~kpc for \(E(k)\). For each method, $\varepsilon$ is calculated by applying the respective theoretical scaling relation (Eqs.~\ref{eq:dll} and \ref{eq:ek}) at its identified characteristic scale, using standard Kolmogorov constants. The corresponding turbulent heating rate per unit volume is then calculated as 
\[
Q_{\rm turb} = \rho \varepsilon,
\]
where $\rho$ denotes the gas density averaged within each radial shell. Fig.~\ref{fig10} plots the heating rates derived independently from the VSF method using $\ell_{\rm VSF}$ and the $E(k)$ method using $\ell_{\rm Ek}$ versus the radiative cooling rates within five clustercentric shells. Since these estimates are based on characteristic scales within the inertial range consistent with Kolmogorov scaling, they should be regarded as order-of-magnitude estimates of the turbulent heating rate rather than precise measurements. As shown in Appendix~\ref{app} (Fig.~\ref{figa1}), the heating rates inferred from the VSF and $E(k)$ methods bracket the thermal energy increase rate measured directly in the simulation. The difference between the two estimates therefore provides a useful indication of the systematic uncertainties associated with the adopted methodology. Throughout this work, we use the VSF- and $E(k)$-based results to characterize a plausible range of turbulent heating rates and to assess the robustness of our conclusions against methodological differences.

Overall, while the two methods yield radial trends of identical shape, the absolute magnitude estimated from $E(k)$ is systematically higher than that derived from the VSF, which may reflect differences in how each method probes the turbulent cascade or sensitivities to anisotropy and compressibility. Despite this quantitative difference, both methods consistently show the same crucial result. From Fig.~\ref{fig10}, it is clear that in the central region(s) the turbulent heating rate derived from either method is significantly smaller than the radiative cooling rate. While the heating rate becomes more comparable to the cooling rate in the outer regions \citep{Walker2018}, the substantial shortfall near the cluster center, where the cooling-flow problem is most severe, indicates that turbulent heating alone is insufficient to balance radiative cooling. This finding agrees with previous simulation studies \citep{Yang16b, Li17, Mohapatra_2019} and suggests that additional heating mechanisms, such as bubble mixing , weak shocks or sound waves \citep{Fabian17,  Wang2022}, are required to prevent catastrophic cooling in cluster cores.

\section{Discussion}\label{sec:4}
\subsection{The Roles of Turbulent Heating in CC Clusters}
Our primary finding, based on the analysis of VSF and \(E(k)\) in our idealized hydrodynamic simulation, shows that the estimated turbulent heating rate falls short of balancing the radiative cooling rate, particularly in the cluster core, indicating that turbulent heating alone cannot offset radiative cooling and highlighting the limitations of turbulence in regulating the energy budget of CC clusters.

To understand the factors affecting this estimate, we examine the properties of the simulated turbulence. In our simulations, both the VSF and $E(k)$ exhibit a well-defined inertial range. The turbulence is driven on large scales (100--250~kpc), allowing a clear cascade to develop below the injection range. This scaling region enables us to estimate the energy transfer rate based on the Kolmogorov hypothesis of a constant energy flux, providing a characteristic measure of the dissipation. This hypothesis, however, is formally valid only for incompressible turbulence, so we must examine the role of compressive motions. Although our pre-existing turbulence is initially driven purely in the solenoidal mode, nonlinear interactions and shock formation naturally generate compressive motions during its evolution \citep{Feder10, Porter15}. These compressive components introduce additional density fluctuations and locally modify the velocity field, which can violate the assumption of incompressibility and produce deviations in both the VSF and \(E(k)\) \citep{Kritsuk07, Feder13, Schmidt09}. Nevertheless, solenoidal motions remain dominant in our simulations, the turbulence within the observed scaling region still follows the expected trends, allowing us to estimate the turbulent heating rate.

Another methodological consideration is the influence of stratification in clusters. According to \citet{Wang_23}, turbulence in the regime with $Fr \gtrsim 0.1$ lies outside the strongly stratified limit and is not strongly suppressed by buoyancy forces. However, as shown in \citet{Mohapatra_2020}, buoyancy can contribute to the redistribution of kinetic energy into potential energy even at these Froude numbers, with this effect increasing continuously with stratification strength. 
%\textcolor{blue}{[KY: Please modify the following sentences to be the same as what's written in the response letter (i.e., most of the regions are within the weak to moderate stratification regime).]} 
Our simulations model a Perseus-like cluster, where the majority of the analyzed volume lies above $Fr \sim 0.35$. This places most of the turbulent gas in a weak to moderate stratification regime. We therefore expect buoyancy effects to be present but not to significantly alter the overall turbulent dynamics relevant to our heating rate estimate.

Despite these complexities, our core conclusion that turbulent heating is insufficient in the core remains broadly consistent with previous simulation studies \citep{Reynolds15, Yang16b, Li17}. This agreement is noteworthy given that our simulations include stronger pre-existing turbulence than many of those studies, suggesting that even enhanced turbulent activity does not provide enough heating to offset cooling in cluster cores. Indeed, invoking turbulent heating as a solution to the cooling-flow problem exhibits some challenges. As discussed in \citet{Fabian17}, if the observed velocity disperson is entirely attributed to turbulence, the turbulent energy would be quickly radiated away and would need to be replenished in order to heat the cluster core. In addition, in contrast to AGN feedback, turbulent heating as driven by mechanisms other than AGNs (e.g., cluster mergers, galaxy motions, plasma instabilities) is not a self-regulated mechanism. While it may contribute to some heating at outer radii  (see Fig.~\ref{fig10}), the excessive cooling near the cluster core would still require other channels of AGN heating.

When comparing to observations, our result appears different from some interpretations of observational data \citep{Zhuravleva14, Zhuravleva16, Zhuravleva2018}, which sometimes infer higher turbulent velocities or suggest that turbulence could play a more dominant heating role. This discrepancy might arise because such analyses often attribute all observed fluctuations to turbulence, while in reality, density fluctuations in the ICM can also be produced by bubbles, contact discontinuities, sound waves, gravity waves, and weak shocks. Consequently, turbulent heating rates inferred from surface brightness fluctuations likely overestimate the true turbulence dissipation rates. %the conversion from observed surface brightness variations to turbulent velocities and heating rates involves inherent uncertainties and strong assumptions.

\begin{figure*}
    \centering
    \includegraphics[width=0.7\textwidth]{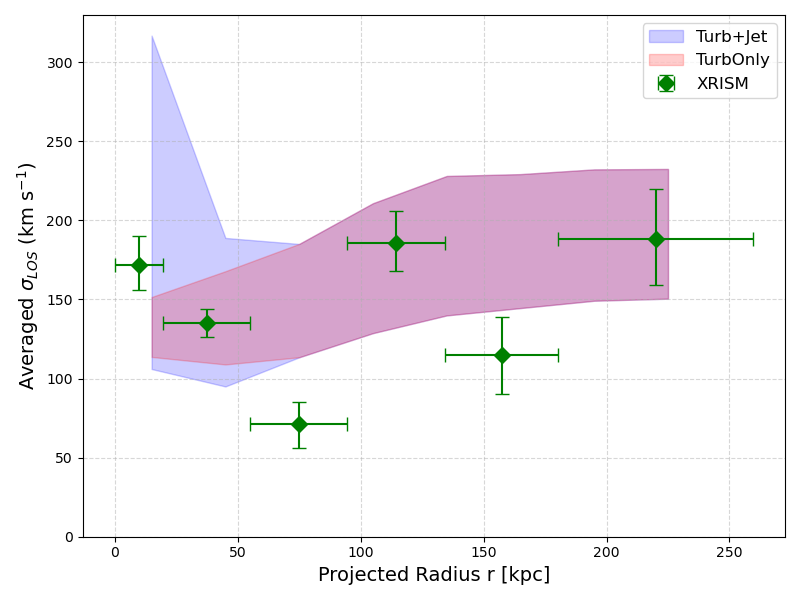}
    \caption{Projected radial profile of the averaged $\sigma_{\rm LOS}$. The shaded regions represent the one-standard-deviation (1$\sigma$) scatter from our \textbf{Turb+Jet} (blue) and \textbf{TurbOnly} (red) simulations, computed 10~Myr after the jet injection. The green data points are the direct \(\sigma_{\rm LOS}\) measurements from the XRISM observation of the Perseus cluster \citep{XRISM25perseus}. Our \textbf{Turb+Jet} simulation successfully reproduces the central enhancement in $\sigma_{\rm LOS}$ observed by XRISM, distinguishing it from the \textbf{TurbOnly} run. This agreement supports our interpretation that while the jet is responsible for the increased central $\sigma_{\rm LOS}$, this is a transient, coherent motion and not a sign of sustained, volume-filling turbulence.
    }
    \label{fig11}
\end{figure*}

\subsection{Interpreting the Velocity Dispersion in Perseus Observed by XRISM}
To examine these observational inferences, resolving the precise contribution of turbulent heating ultimately requires direct measurements of ICM velocity structures. The high spectral resolution X-ray capabilities of the recently launched X-Ray Imaging and Spectroscopy Mission (XRISM) mission are beginning to provide such constraints through measurements of turbulent line broadening. Their initial results, consistent with the earlier landmark findings from \textit{Hitomi} for the Perseus cluster core \citep{Hitomi16}, confirm that the gas motions are relatively quiescent, with $\sigma_{\rm LOS} \lesssim 200~\mathrm{km\,s^{-1}}$ \citep{XRISM25perseus, XRISM25A2029b, XRISM25centaurus}. These low velocities imply substantially less turbulent kinetic energy than the amount required to fully offset the radiative cooling losses in the cluster core. A similar conclusion is drawn from the Hydra A cluster \citep{Rose2025}, which reports a low $\sigma_{\rm LOS} \approx 164~\mathrm{km\,s^{-1}}$. They find that the total turbulent dissipation is nearly an order of magnitude lower than the total radiative cooling luminosity ($L_{\rm cool}/L_{\rm turb} \approx 6$). This comparison with Hydra A observations further supports our interpretation that jet-driven gas motions alone are unlikely to maintain thermal balance through stochastic turbulence.

Beyond just the total kinetic energy, the ability of turbulence to redistribute energy throughout the CC is limited by the efficiency of turbulent diffusion. The effective turbulent diffusion coefficient can be estimated as $D_{\rm turb} \sim C_\mu \, k_{\rm turb}^2 / \varepsilon$, where $k_{\rm turb}$ is the turbulent kinetic energy per unit mass and $\varepsilon$ is the corresponding dissipation rate. Adopting $C_\mu \approx 0.09$ and using characteristic values from our simulations, this yields $D_{\rm turb} \sim 7 \times 10^{29}~\mathrm{cm^2\,s^{-1}}$, consistent with the estimates reported by \citet{Vazza12}. This value implies diffusion timescales that are comparable to or longer than the cooling time in the cluster core, suggesting that turbulence is unlikely to efficiently propagate and distribute heat across the entire cool core \citep{Fabian17, Rose2025}.

However, \cite{XRISM25perseus} also reports a new detail in the Perseus cluster, showing that $\sigma_{\rm LOS}$ gradually increases toward the center within the inner 60~kpc. They interpret this central enhancement as direct evidence of ``AGN-powered turbulence.'' Based on this interpretation, they suggest that this turbulence, if fully dissipated, could play a significant role in offsetting radiative cooling losses in the core. Our simulations provide a direct test of this interpretation. Specifically, the simulated $\sigma_{\rm LOS}$ profiles shown in Fig.~\ref{fig11} were computed by projecting the 3D velocity field along the LOS, followed by radial binning and azimuthal averaging to enable a clean comparison with the observational data. As shown in Fig.~\ref{fig11}, our \textbf{Turb+Jet} simulation successfully reproduces this central increase in $\sigma_{\rm LOS}$, distinguishing it from \textbf{TurbOnly} and remaining consistent with the XRISM data points. This apparent agreement, however, highlights a critical distinction in interpretation. As our analysis in Section~\ref{sec:3.4} demonstrated, while these jet-driven gas motions are responsible for the observed increase in central $\sigma_{\rm LOS}$, they are fundamentally transient, unresolved bulk flows that do not effectively feed the stochastic turbulent cascade. Therefore, our results contradict the view that this central enhancement represents sustained turbulence, suggesting that the turbulent heating rate may be overestimated if the observed velocity dispersion near the center is interpreted as turbulent motions driven by AGN jets. This aligns with the caveat noted by \cite{XRISM25perseus} that their turbulent heating rate estimate could be overestimated if the measured $\sigma_{\rm LOS}$ was significantly driven by unresolved coherent bulk motions.

\subsection{Limitations of Current Simulations}
In this work, we investigate the fundamental properties and evolution of turbulence in the ICM using an idealized hydrodynamic setup. We have considered only a single injection of AGN energy, which contributes little to the overall velocity field and turbulent energy in the ICM. In contrast, self-regulated AGN feedback injects energy quasi-continuously, generating more sustained turbulence over longer timescales and larger spatial scales. How such self-regulated activity interacts with pre-existing turbulence remains uncertain.

To isolate the evolution of turbulence, we adopt an idealized hydrodynamic setup that excludes additional physical processes such as magnetic fields, radiative cooling, and thermal conduction. This allows us to study how turbulence develops, cascades, and dissipates under controlled conditions, including its interaction with an impulsive AGN outburst. However, these neglected processes can substantially influence both the driving and dissipation of turbulence in realistic cluster environments. Radiative cooling in cluster cores drives thermal instabilities, forming dense, cold gas and multiphase structures \citep{McCourt12, Sharma12, Gaspari13}. Turbulence can interact with this cold phase, enhancing mixing and modifying the velocity field and energy distribution \citep{Gaspari18, Li14, Wang_21}. Magnetic fields, meanwhile, can be amplified by turbulence via a dynamo, converting kinetic into magnetic energy \citep{Feder16, Di21}, and introduce anisotropy that resists motions perpendicular to field lines, thereby modifying the turbulent cascade, the slope of VSF \citep{Mohapatra2022a}, and vorticity evolution \citep{Porter15}. They also provide pressure support and influence plasma instabilities in the weakly collisional ICM \citep{Santos14, Kunz14, Kingsland19}. Thermal conduction, which is highly anisotropic along field lines in the ICM \citep{Narayan01}, can smooth temperature gradients, reduce the local requirement for turbulent heating \citep{Zakamska03, Parrish08, Yang16a}, and preferentially damp small-scale fluctuations, affecting the shape of \(E(k)\) and VSF in the dissipation range \citep{Ruszkowski10}. Omitting these processes simplifies the thermal and dynamical structure, removing important pathways that shape turbulence and its energy transport in the ICM.

In our simulations, energy dissipation occurs at the grid scale, which is physically justified by observations showing that the ICM’s viscosity is strongly suppressed \citep{Zhuravleva19, Kingsland19}. This grid-scale dissipation provides a reasonable approximation for how turbulence decays in the high-Reynolds-number ICM and is expected to have minimal impact on the overall energy budget.

Finally, our simulations neglect cosmic rays (CRs), expected from AGN activity \citep{Drury83}. CRs contribute non-thermal pressure, affecting dynamics such as bubble evolution \citep{Yang19, Lin23}, and can directly heat the ICM via plasma processes like the streaming instability \citep{Wiener13, R17}, providing a heating channel distinct from hydrodynamic turbulence. The complex coupling between CRs, magnetic fields, and turbulence can alter the cascade dynamics and effective dissipation rate. Neglecting CRs therefore simplifies the energy budget, potentially affecting our assessment of turbulent heating relative to radiative cooling.

We will address some of these limitations in future work by including additional physics and more realistic feedback modes, providing a more complete understanding of turbulence and its contribution to the thermal balance in cluster cores.

\section{Conclusions}\label{sec:5}
In this study, we performed 3D hydrodynamic simulations to investigate the role of turbulent heating in a Perseus-like cluster. We considered three simulation setups: one with only pre-existing turbulence (\textbf{TurbOnly}), one with only the AGN jet (\textbf{JetOnly}), and one including both (\textbf{Turb+Jet}). The forcing amplitude of the pre-existing turbulence in \textbf{TurbOnly} and \textbf{Turb+Jet} was calibrated to match the observationally constrained $\sigma_{\rm LOS}$ in the cluster core. These controlled setups allow us to assess the individual and combined contributions of turbulence and AGN feedback to ICM heating. Our primary goal was to determine whether turbulent heating could balance radiative cooling, focusing on resolving discrepancies between simulation results and observations. We estimated the turbulent dissipation rate $\varepsilon$ using two independent methods based on Kolmogorov theory by analyzing the VSF and energy power spectrum.
Our key findings can be summarized as follows:
\begin{itemize}
\item In our simulations, we find that the single AGN jet injection has a minimal impact on the global turbulent velocity statistics (e.g., $\sigma_{\rm LOS}$), which remains dominated by the pre-existing turbulence. However, the jet still clearly deposits energy into the ICM, as evidenced by the increase in the entropy profiles and the presence of jet-driven pressure perturbations ($\Delta P/P$) associated with weak shocks and sound waves. This indicates that, despite its inability to stir the global turbulent velocity field, the jet can still heat the ICM through bubbles, weak shocks, and sound waves.

\item Our analysis of the early-time $E(k)$ suggests that a single AGN jet injection is a relatively inefficient driver of sustained turbulence. The energy it injects is transient and does not cascade down to smaller scales. Instead, the jet-driven motions are best described as short-lived, large-scale coherent bulk motions, leaving the pre-existing turbulence to dominate the cluster core's turbulent energy budget.

\item The turbulence in our simulations, characterized by analyses of $D_{\rm LL}$ and $E(k)$, shows a well-defined inertial range. Although the flow also includes compressive modes (despite predominantly solenoidal driving), it remains largely dominated by solenoidal motions and resides in a weak to moderate stratification regime, supporting the approximate applicability of Kolmogorov-based cascade estimates despite systematic methodological uncertainties.

\item We estimate $\varepsilon$ by applying the respective scaling relations (Eqs.~\ref{eq:dll} and \ref{eq:ek}) at characteristic scales within the observed inertial range ($\ell_{VSF} \approx 53.1$~kpc and $\ell_{Ek} \approx 52.6$~kpc). While both methods show similar radial trends, the heating rate derived from $E(k)$ is systematically higher than that from $D_{\rm LL}$. Despite this quantitative difference, both methods consistently conclude that the turbulent heating rate is insufficient to offset the radiative cooling rate within the cluster core (r $<$ ~50 kpc), although they become more comparable at larger radii.

\item Our conclusion that turbulent heating alone is subdominant in the core aligns with previous simulation studies that focused solely on AGN-driven turbulence \citep{Reynolds15, Yang16b, Li17, Mohapatra_2019}. This suggests that even when observationally constrained levels of pre-existing turbulence \citep{Hitomi16} are included, as in our work, turbulent heating alone may still be insufficient to solve the cooling-flow problem.

\item Our results highlight a tension with interpretations of some observational data based on X-ray surface brightness fluctuations  \citep{Zhuravleva14, Zhuravleva16}, which sometimes infer higher turbulent velocities and suggest a dominant role for turbulent heating. This discrepancy potentially stems from the assumptions required to infer turbulent velocities from density fluctuations.

\item Our conclusion of insufficient turbulent heating is broadly supported by direct \(\sigma_{\rm LOS}\) measurements from \textit{Hitomi} and XRISM, which show overall quiescent gas \citep{Hitomi16, XRISM25perseus, XRISM25A2029b, XRISM25centaurus}. However, we challenge the interpretation of the central \(\sigma_{\rm LOS}\) enhancement reported in Perseus \citep{XRISM25perseus}. Our analysis demonstrates that the enhanced $\sigma_{\rm LOS}$ near the cluster center is due to transient, unresolved coherent bulk motions driven by the jets, rather sustained turbulence. Interpreting the enhanced $\sigma_{\rm LOS}$ as turbulence would overestimate the true turbulent heating rate.

\end{itemize}
Therefore, based on our hydrodynamic simulations, we conclude that turbulent heating, even when considering pre-existing turbulence interacting with AGN feedback, is unlikely to be the primary heating mechanism balancing radiative cooling in the central regions of CC clusters like Perseus. While turbulence undoubtedly plays a role in the ICM's dynamics and energy transport, other heating mechanisms (e.g., AGN-driven bubble mixing, weak shocks, sound waves, CR heating) appear necessary to fully address the cooling-flow problem. Future work incorporating more comprehensive physics, including radiative cooling, magnetic fields, CRs, thermal conduction, and self-regulated feedback cycles, will be crucial for a complete understanding of the ICM energy budget.

%% Please use the acknowledgment and contribution environments. This will 
%% be anonomyized when the "anonymous" style option is used. 
\begin{acknowledgments}
JLL and HYKY acknowledge support from National Science and Technology Council (NSTC) of Taiwan (NSTC 112-2628-M-007-003-MY3; NSTC 114-2112-M-007-032-MY3). HYKY acknowledges support from Yushan Scholar Program of the Ministry of Education (MoE) of Taiwan (MOE-108- YSFMS-0002-003-P1). This work used high-performance computing facilities operated by Center for Informatics and Computation in Astronomy (CICA) at National Tsing Hua University (NTHU). This equipment was funded by the MoE of Taiwan, the NSTC of Taiwan, and NTHU. FLASH was developed in part by the DOE NASA- and DOE Office of Science-supported Flash Center for Computational Science at the University of Chicago and the University of Rochester.

\end{acknowledgments}

\section*{Data Availability}
\noindent The data underlying this article will be shared upon reasonable request to the corresponding author. The simulations were performed using the publicly available FLASH code. The AGN feedback module used in this work is a proprietary extension of the FLASH framework and is therefore not publicly available. All other components of the simulation setup are based on standard FLASH implementations and established methods described in the literature.

%The data underlying this article will be shared upon reasonable request to the corresponding author. \textcolor{blue}{[KY: Please update this section according to the reply to the Data Editor.]}

\appendix
\section{Validation of the turbulent heating rate}\label{app}
\begin{figure*}
    \centering
    \includegraphics[width=0.8\textwidth]{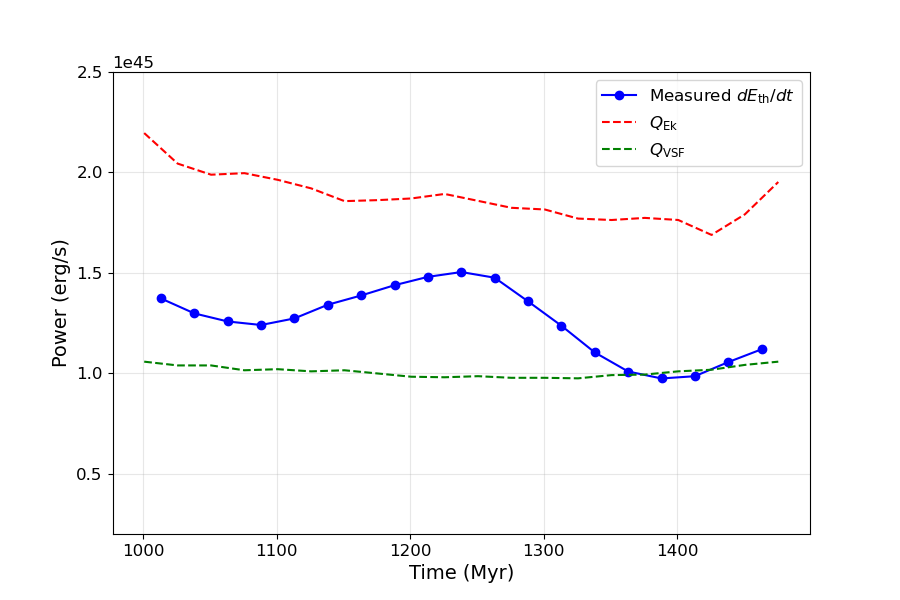}
    \caption{Validation of the turbulent heating rate for the \textbf{TurbOnly} simulation. The red dashed line shows the dissipation rate inferred from the energy spectrum ($Q_{\rm Ek} = \varepsilon_{\rm Ek} \cdot M_{\rm total}$), and the green dashed line shows the corresponding estimate from the velocity structure function ($Q_{\rm VSF} = \varepsilon_{\rm VSF} \cdot M_{\rm total}$). The blue points represent the thermal energy increase rate ($dE_{\rm th}/dt$) measured directly from the simulation. All quantities are shown as a function of time.
    }
    \label{figa1}
\end{figure*}

To validate the turbulent dissipation rate inferred from the velocity structure function and energy spectrum methods, we compare these estimates with the direct measurement of the thermal energy increase rate for the \textbf{TurbOnly} simulation. As shown in Fig.~\ref{figa1}, both estimators are consistent with the thermal energy increase rate measured directly from the simulation. The direct measurement provides an independent reference for the actual conversion rate of kinetic energy into thermal energy in the simulated ICM. The agreement within within a factor of $\sim 2$ indicates that the two statistical methods capture the correct order of magnitude of turbulent dissipation. The remaining differences between the two estimators reflect systematic uncertainties associated with scale selection, anisotropy, and numerical dissipation effects inherent in the two approaches.

\restartappendixnumbering

\bibliography{agn}
\bibliographystyle{aasjournalv7}



\end{document}